\documentclass[12pt]{article}
\usepackage{amsmath,amssymb}
\usepackage{graphicx}
\usepackage{subfig}

\newsavebox{\uuunit}
\sbox{\uuunit}
    {\setlength{\unitlength}{0.825em}
     \begin{picture}(0.6,0.7)
        \thinlines
        \put(0,0){\line(1,0){0.5}}
        \put(0.15,0){\line(0,1){0.7}}
        \put(0.35,0){\line(0,1){0.8}}
       \multiput(0.3,0.8)(-0.04,-0.02){10}{\rule{0.5pt}{0.5pt}}
     \end {picture}}

\allowdisplaybreaks

\numberwithin{equation}{section}
\begin{document}
\begin{titlepage}
\begin{center}
%\vskip 6mm
\vskip 3in
%%%%%%%%%%%%%%%%%%%%%%%%%%%%%%%%%%%%%%%%%%
{\Large \textbf{A comparative study of $2d$ Ising model at different boundary conditions using Cellular Automata}}
%%%%%%%%%%%%%%%%%%%%%%%%%%%%%%%%%%%%%%%%%%%
\vskip 8mm
{$^{\dag}$Jahangir Mohammed\footnote{
Present address: Department of Physics, Nabarangpur College, Nabarangpur-764063, Odisha, India.} and $^{*,\ddag}$Swapna Mahapatra}

{\em Department of Physics, Utkal University,\\
Bhubaneswar, Odisha 751004,
India\\
$^{\dag}$jahangirmd.physics@gmail.com\\
$^{*}$swapna.mahapatra@gmail.com\\
$^{\ddag}$swapna@iopb.res.in}
%
%{\tt jahangir.isi@gmail.com}\;,\;\, {\tt swapna@iopb.res.in}
\end{center}

\vskip .2in
%%%%%%%%%%%%%%%%%%%%%%%%%%%%%%%%%%%%%%%%%%%%%%%%%%%%%
\begin{center} {\bf Abstract } \end{center}
\begin{quotation}\noindent
Using Cellular Automata,
we simulate spin systems corresponding to $2d$
Ising model with various kinds of boundary conditions
(bcs). The appearance of spontaneous magnetization
in the absence of magnetic field is studied with a
$64\times64$ square lattice with five different bcs, i.e.,
periodic, adiabatic, reflexive, fixed ($+1$ or $-1$)
bcs with three initial conditions (all spins up, all
spins down and random orientation of spins). In the
context of $2d$ Ising model, we have calculated
the magnetisation, energy, specific heat, susceptibility and
entropy with each of the bcs and observed that the
phase transition occurs around $T_c$ = 2.269 as obtained by
Onsager. We compare the behaviour of magnetisation vs
temperature for different types of bcs by calculating the
number of points close to the line of zero magnetisation
after $T>T_c$ at various lattice sizes. We
observe that the periodic, adiabatic and reflexive bcs give
closer approximation to the  value of $T_c$ than fixed +1
and fixed -1 bcs with all three initial conditions
for lattice size less than $70\times70$. However, for lattice size
between $70\times70$ and $100\times100$, fixed +1 bc and
fixed -1 bc give closer approximation to the $T_c$ with
initial conditions all spin down configuration and all spin up
configuration respectively.
%%%%%%%%%%%%%%%%%%%%%%%%%%%%%%%%%%%%%%%%%%%%%%%%%%%%%%%%%%%%%%%%%%%%
\end{quotation}
\vfill
%%%%%%%%%%
%\today
%%%%%%%%%%%%%%%%%%
\end{titlepage}
%%%%%%%%%%%%%%%%
\eject
%%%%%%%%%%%%%%%%
%%%%%%%%%%%%%%%%%%%%%%%%%%%%%%%%%%%%%%%%%%%%%%%%%%%%%%%%
%%%%%%%%%%%%%%%%%%%%%%%%%%%%%%%%%%%%%%%%%%%%%%%%%%%%%%%%
\section{Introduction}
\label{sec:introduction}
\setcounter{equation}{0}
%%%%%%%%%%%%%%%%%%%%%%%%%%%%%%%%%%%%%%%%%%%%%%%%%%%%%%%%
%%%%%%%%%%%%%%%%%%%%%%%%%%%%%%%%%%%%%%%%%%%%%%%%%%%%%%%%
The phenomenon of magnetism belongs to one of the oldest observations
in nature which is yet to be understood at a fundamental level. One
remarkable effect is the appearance of spontaneous magnetization
giving rise to ferromagnetism when certain materials  are cooled
down below a critical temperature called Curie temperature in the
absence of any external applied magnetic field. The $2d$ Ising model is
represented by a square lattice of particles, each carrying one of
the two spins states with magnetic moments $\pm1$. Each particle at
a node is assigned a definite orientation. Spins of these particles
cause a magnetic field whose strength decreases with increase in
distance in the lattice. For simplification, we consider only the nearest
neighbour interaction i.e., no other particle is located closer to one of them.
In $2d$ Ising model, an ordinary particle has four nearest neighbours at east,
west, north and south direction of the particle. These spin interactions
contribute to the energy of the whole system. The energy of a spin configuration
$s = \{s_{i,j}, s_{i,j} \in \{+1, -1\},i,j=1, \dots, L \}$, with $L$ as the order of
square lattice is given by the Hamiltonian

\begin{equation}
\label{eq:eqn3}
 H(s) = - \sum\limits_{i,j=1}^{L} J_{i,j} s_{i,j} (s_{i,j-1}+s_{i,j+1}+s_{i-1,j}+s_{i+1,j})
 - \mu \sum\limits_{kl} H_{k,l} s_{k,l}
\end{equation}
Where $J_{i,j}$ is the  exchange interaction among $s_{i,j}$ with their four neighbours,
$\mu$ is the magnetic moment and  $H_{k,l}$ is the external magnetic field at
$(k,l)^{th}$ spin. For simulation purpose, we have to define a finite system
with $L^2<\infty$. We study different bcs under which the interaction energy will
be  maximum. In a periodic bc, the matrix $J_{i,j}$ defines a nearest neighbourhood
topology of a loop and for other bcs nearest neighborhood topology is a square of
a square lattice. A $2d$ Ising model with $L^2$ particles has $2^{L^2}$ spin
configurations.

The  partition function in Boltzmann statistics is given by

 \begin{eqnarray}
 \label{eq:eqn4}
 Z_\beta = \sum\limits_{s} e^{-\beta H(s)}
\end{eqnarray}
\noindent
where $ \beta = \frac{1}{k_B T}$, where $k_B$ is Boltzmann constant.
If we consider a $8\times8$ lattice, the space
of states $s$ has $2^{64}$ elements and it is a daunting task to compute
$Z_{\beta}$. To find a concise formula for $Z_\beta$, the thermodynamic
limit $L \rightarrow \infty$ is considered in analytical calculation.
Basing on the transfer matrix method with pbc, Onsager has solved the
$2d$ Ising model \cite{1}. Kotecky et. al \cite{2} have studied
magnetization of the Ising model under minus fixed bc. Still the $2d$
Ising model with other bcs are yet to be solved. So, here we consider
five boundary conditions (bcs) to simulate $2d$ Ising model.

Explicit formulation of the spontaneous magnetization of the $2d$ Ising model with
pbc for $L \rightarrow \infty$ was carried out in reference \cite{3}
and the magnetization $m_\beta$ in terms $\beta$ is found to be,
%\begin{eqnarray}
\begin{displaymath}
   m_\beta = \left\{
     \begin{array}{lr}
       (1-(\sinh(2J\beta))^{-4})^{1/8}, & \beta > \beta_c\\
       0, & \beta \leq \beta_c
     \end{array}
   \right.
\end{displaymath}
%\end{eqnarray}
where
\begin{eqnarray*}
 \beta_c &=& \frac{\log (1 + \sqrt{2})}{2J}  \\
 \Rightarrow  T_c &=& 2.269 \quad (or \, exact \, solution).
\end{eqnarray*}

where $k_B = 1$ and $J = 1$ for a ferromagnetic substance.

Magnetisation in terms of $T$ and $T_c$ is given by,

\begin{equation}
\label{eq:eqn5}
   m = \left\{
     \begin{array}{lr}
       (1-(\sinh(\log(1+\sqrt{2})\frac{T_c}{T})^{-4})^{1/8}, & T < T_c\\
       0, & T \geq T_c
     \end{array}
   \right.
\end{equation}

From \emph{equation 3}, it is seen that magnetisation has at least two
different possible directions and the average magnetization is zero in
the absence of external magnetic  field at $T > T_c$. We consider the
above theory to compare among different bcs.

Cellular Automaton is a mathematical model in which the state of a cell
interact with neighbours and then updates the state according to a
specific rule in $2d$ CA \cite{4}. This transition rule depends
on the problem on which one is interested. While dealing with different
dimensions, CA models are categorised as $1D$, $2D$, $3D$ CA etc.
In $2D$ CA, cells may be square, triangular, hexagonal, polygon type.
State of the cell is given  in terms of any finite number.
The  number of neighbours depend on the dimension and the specific
approach to the problem. To simulate the Ising model, we can create
a $2$ states CA, for spin up state ($+1$) and spin down state ($-1$).
For $1D$ model, we can consider two (nearest neighbours) or
four neighborhoods (next nearest neighbours), for $2D$
we can consider four (north, south, east and west),
six (honeycomb lattice) or eight neighborhoods (north, south, east,
west and four corners neighbours)and for $3D$ we can consider six (north, south,
east, west, top and bottom) or twenty six neighborhoods (north, south,
east, west, top, bottom and twenty corners neighbours)\cite{4}. Both the CA
model and the Ising model have similar characteristics. However, in
Ising model case, before $T_c$ states of the cells are either all in up
state or all in down state and after $T_c$, the net magnetisation becomes
zero and the pattern become random (on the average half in $+1$ spin states
and half in $-1$ spin states). So, it is a big challenge to find a
specific rule in CA that satisfies the above behaviour of the Ising model.

Numerical methods like Markov chain, Metropolis \cite{5},
Wolff algorithm \cite{6} take a lot of time to simulate the Ising model.
Monte Carlo is one of the simulation methods which has been widely used for
studying Ising models \cite{7}. Lot of work has been done for mapping
Ising models using CA. A deterministic CA (DCA) is mostly used for this purpose.
Domany and Kinzel \cite{8} modelled a DCA in triangular lattice with
conditional probabilities as the transition rule which maps $2d$ Ising model by
the directed percolation. The so called
Q2R CA \cite{9} (so named by G\'{e}rard Vichniac \cite{10})
is a deterministic, reversible, nonergodic and fast method
that is used for the microcanonical Ising model. Many authors have
produced results based on this model \cite{11,12,13}.
The Creutz CA \cite{14} has simulated the $2d$ Ising model successfully
near the critical region under periodic bc and using this Creutz CA, the
Ising model simulations in higher dimensions e.g., in
$3D$  \cite{15}, $4D$  \cite{16}, $8D$ \cite{17} have been done.
Although the Q2R and Creutz CA models are deterministic and fast, it has been
demonstrated that the probabilistic model of the CA like Metropolis algorithm
is more realistic for description of the Ising model even though the random
number generation makes it slower. Probabilistic CA model under periodic bc
has been studied in the context of an anisotropic-layer (nearest-neighbor
interactions within each layer are different) Ising and Potts models
to find the critical point and shift exponent between two-layers \cite{18}.
However, Ising model using two dimensional CA under
different bcs other than period has not yet been studied.

The paper is organised as follows: in \emph{section 2}, we discuss the basic
theory to treat a $2D$ CA and how to implement
it in the Ising model. The simulation result and discussions are given in
\emph{section 3}. The  comparison of the  five bcs with three different
initial conditions are discussed in \emph{section 4}. Our conclusion and
future perspective are discussed in \emph{section 5}.

%%%%%%%%%%%%%%%%%%%%%%%%%%%%%%%%%%%%%%%%%%%%%%%%%%%%%%%%%

\section{Implemetation of Isotropic $2d$ Ising Model by Square-Lattice CA}
\label{sec:cellular-automata}
%\label{sec:deformed-special}

\setcounter{equation}{0}
%%%%%%%%%%%%%%%%%%%%%%%%%%%%%%%%%%%%%%%%%%%%%%%%%%%%%%%%%

Two dimensional CA is described by finite states of cells ($s$),
neighborhood cells ($n$) and its distance among neighbourhood
($r$), boundary conditions and transition functions or
rules ($f$). In our 2D CA model, $s = \{s_{i,j},\, s_{i,j}\in-1/+1\}$,
number of neighbours $n = 4$ (four nearest neighbours),
$r = 1$ and we consider all five bcs.

Neighbourhoods of extreme cells are taken care of by bc.
In fixed bc, the extreme cells are connected to $-1$ or
$+1$ state. If it is connected to $+1$ state, it is called
fixed $+1$ bc (f1bc) and if it is connected to $-1$ state,
then it is called fixed -1 bc (f-1bc). If the extreme cells
are adjacent to each other then it is called periodic bc (pbc).
In adiabatic bc (abc), the extreme cells replicate their state
and in reflexive bc (rbc), mirror position states replace the extreme cells.

If the same rule is applied to all the elements of the matrix,
then it is called \emph{uniform CA} and if different rules are
applied to all the elements of the matrix or block of elements
then it is called \emph{nonuniform CA}. At different time
intervals, if different rules are applied to the matrix then it is
called varying CA e.g., \emph{probabilistic CA}. With the
application of these rules, elements (states) of the matrix
change at successive intervals as shown in the following equation.

\begin{eqnarray}
\label{eq:eqn13}
s_{L \times L}^{t+1} = f_{L \times L}^{t} \times s_{L \times L}^{t}
\end{eqnarray}
where $f$ is a time varying rule or transition matrix.

Consider an isotropic $2d$ Ising Model in the form of square lattice
($s$) with $L$ rows and $L$ columns.
Lattice has then $L^2=N$ sites. Each of the site $s_{i,j},\,i,j=1,\,\dots,\,L$
in such a way that $i$ increases from left to right and $j$ increases
from top to bottom, has one of the $\pm 1$ spin, which are two states in CA. So,
there are $2^{L^2}$ spin configurations. We consider the nearest
neighbor interactions, so the number of neighbor is $4$. We include the
five different bcs as
\begin{enumerate}
 \item pbc : $s_{i,L+1} = s_{i,1}$, $s_{L+1,j} = s_{1,j}$,\\
             $s_{i,0} = s_{i,L}$ and $s_{0,j} = s_{L,j}$.
 \item abc : $s_{i,L+1} = s_{i,L}$, $s_{L+1,j} = s_{L,j}$,\\
             $s_{i,0} = s_{i,1}$ and $s_{0,j} = s_{1,j}$.
 \item rbc : $s_{i,L+1} = s_{i,L-1}$, $s_{L+1,j} = s_{L-1,j}$,\\
             $s_{i,0} = s_{i,2}$ and $s_{0,j} = s_{2,j}$.
 \item f1bc : $s_{i,L+1} = +1$, $s_{L+1,j} = +1$,\\
              $s_{i,0} = +1$ and $s_{0,j} = +1$.
 \item f-1bc : $s_{i,L+1} = -1$, $s_{L+1,j} = -1$,\\
               $s_{i,0} = -1$ and $s_{0,j} = -1$.
\end{enumerate}

Average magnetization for the configuration is defined as,

\begin{equation}
\label{eq:eqn23}
 \left\langle M \right\rangle = \sum\limits_{i,j=1}^{L} s_{ij}
\end{equation}

and the average magnetization per spin is given by,

\begin{equation}
\label{eq:eqn24}
\left\langle m \right\rangle = \frac{\left\langle M \right\rangle}{N}
\end{equation}

Energy for the configuration $s$ is defined as,

\begin{equation}
\label{eq:eqn25}
 E(s) = - \frac{J}{2}\sum\limits_{i,j=1}^{L} s_{ij} \times (s_{i-1,j}+s_{i+1,j}+s_{i,j-1}+s_{i,j+1})
\end{equation}

Here, the factor of $1/2$ has been put to remove the double
counting of energy otherwise the interacting energy will be computed twice.
$J_{ij} = J$ (isotropic) for $4$ neighbours, or else, $J_{ij}=0$.

The configuration energy per spin is

\begin{equation}
\label{eq:eqn26}
\left\langle e \right\rangle  = \frac{E(s)}{N}
\end{equation}

For updating the lattice in next iteration, we use the probabilistic
approach by constructing a probabilistic CA. We use the following procedure.

First we calculate the change in energy, $\Delta E(s^t) = E(s^t) - E(s^{t-1})$
i.e., the energy difference at successive time intervals. We do not consider
the case when $\Delta E < 0$, because it is obvious that after a finite time,
system falls to ground state, and there can not be a state with lower energy.
In our approach, we consider $E(s^t) \ge E(s^{t-1})$. Next we calculate
the probability of each site in the spin configuration $s$ at time $t$
(number of iterations) by using the Boltzmann factor

\begin{equation}
\label{eq:eqn27}
 p_t = \frac{p(E(s^t))}{p(E(s^{t-1}))} = e^{-\frac{\Delta E(s^t)}{k_BT}}
\end{equation}

With the above probability for each site, we construct a probability weighted matrix
(or transition matrix). This matrix leads to our probabilistic CA matrix ($PCA^t$)
by comparing with a random matrix and multiplying by a factor $0.1$
to normalise the $PCA^t$.

Successive spin configurations are obtained from

\begin{equation}
\label{eq:eqn28}
 [s_{i,j}^{t+1}]_{L\times L} = [PCA_{i,j}^t]_{L\times L} [s_{i,j}^{t}]_{L\times L}
\end{equation}

After a finite iteration we calculate the average energy of the
system per site ($e$), magnetisation per site ($m$), susceptibility per site ($\chi$),
specific heat per site ($C_v$) and entropy ($S$), where,

\begin{equation}
\label{eq:eqn29}
 \chi = \frac{N}{k_B T}(\left\langle m^2 \right\rangle -
 \left\langle m \right\rangle ^2)
\end{equation}

\begin{equation}
\label{eq:eqn30}
 C_v = \frac{N}{k_B T^2}(\left\langle E^2 \right\rangle -
 \left\langle E \right\rangle^2)
\end{equation}

\begin{equation}
\label{eq:eqn31}
 S = -k_B (r_1 P_1 \log_2 P_1 - r_2 P_2 \log_2 P_2)
\end{equation}

Where $r_1$ is the total number of spin up states, $r_2$ is the total number
of spin down states, $P_1$ is the probability of spin up states and $P_2$
is the probability of spin down states in the lattice $s$. Our
probabilistic CA matrix updates in successive time and every spin
that is updated in the direction of higher energy will be unflipped
in the next iteration. This algorithm checks the time complexity better
than the Metropolis algorithm \cite{5} that transits one
spin at a time.

%%%%%%%%%%%%%%%%%%%%%%%%%%%%%%%%%%%%%%%%%%%%%%%%%%%%%%%%%%%%%
\section{Simulation Results and Discussions}
\label{sec:simulation}
%%%%%%%%%%%%%%%%%%%%%%%%%%%%%%%%%%%%%%%%%%%%%%%%%%%%%%%%%%%%%
In this work, we have considered square lattice of different
sizes with $J = 1$ and $k_B = 1$. We do not consider external
magnetic field $H$. Here, temperature $T$ ranges from $0.1$
to $5.0$ (as we study the phase transition). We carry out
the simulation with all the three initial conditions and with
all five bcs. Here three initial conditions are (i) all up
(or most of spins up/$+1$), (ii) all down (or most of the
spins down/$-1$) and (iii) random (or randomly oriented spins up/down).
The optimal lattice size and maximum iteration are decided
by the simulation result, which is relevant to study the
phase transition.

%%%%%%%%%%%%%%%%%%%%%%%%%%%%%%%%%%%%%%%%%%%%%%%%%%%%%%%%%%%%%
\subsection{Simulation to find maximum iteration}
\begin{figure}[ph]
\centerline{\includegraphics[width= 13.0cm]{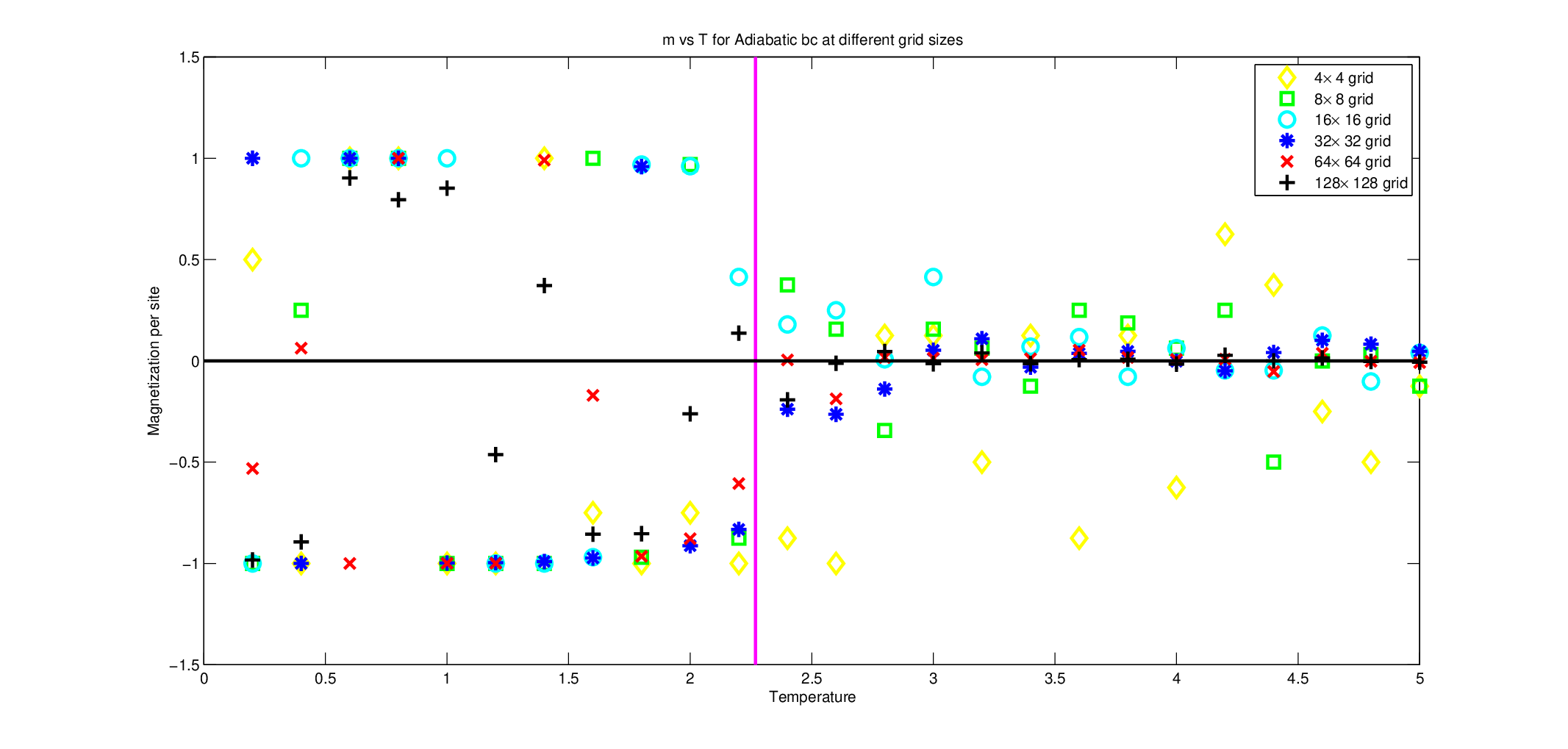}}
\vspace*{8pt}
\caption{Magnetisation per site vs temperature for the
comparison among lattice sizes ($4\times4$,
$8\times8$, $16\times16$, $32\times32$, $64\times64$
and $128\times128$) with randomly oriented spin
configuration as initial condition and abc. Magenta line represents a
parallel line to magnetisation per site at $T=T_c$. \label{f1}}
\end{figure}

%%%%%%%%%%%%%%%%%%%%%%%%%%%%%%%%%%%%%%%%%%%%%%%%%%%%%%%%%%%%%
In this simulation, we have found the maximum iteration time
($t_{max}$) by applying our transition rule to compare between
different bcs and for calculation of magnetisation per site $\bf m$,
energy per site $\bf e$, $\bf \chi$, $\bf C_v$ and $\bf S$. We have done all the
above calculations by taking the average of ten simulations. For
our initial guess of $t_{max}=2^{15}$ (for optimization purpose
and to avoid the time complexity), we have considered lattice sizes $4\times4$,
$8\times8$, $16\times16$, $32\times32$, $64\times64$ and $128\times128$
with all three initial conditions and all the five bcs. In figure ~\ref{f1},
with abc, $64\times64$ and $128\times128$ lattice sizes show $m \approx 0$
after $T > T_c$ i.e., magnetisation fluctuates less around zero value of magnetisation.
After several runs on different bcs with all three initial conditions,
we have found $t_{max}$, by taking lattice sizes $64\times64$ and $128\times128$.
Here, we have chosen $64\times64$ lattice size for reducing the computational time.

%%%%%%%%%%%%%%%%%%%%%%%%%%%%%%%%%%%%%%%%%%%%%%%%%%%%%%%%%%%%%%%%%%%%
%%%%%%%%%%%%%%%%%%%%%%%%%%%%%%%%%%%%%%%%%%%%%%%%%%%%%%%%%%%%%%%%%

\begin{figure}[ph]
\centerline{\includegraphics[width= 13.0cm]{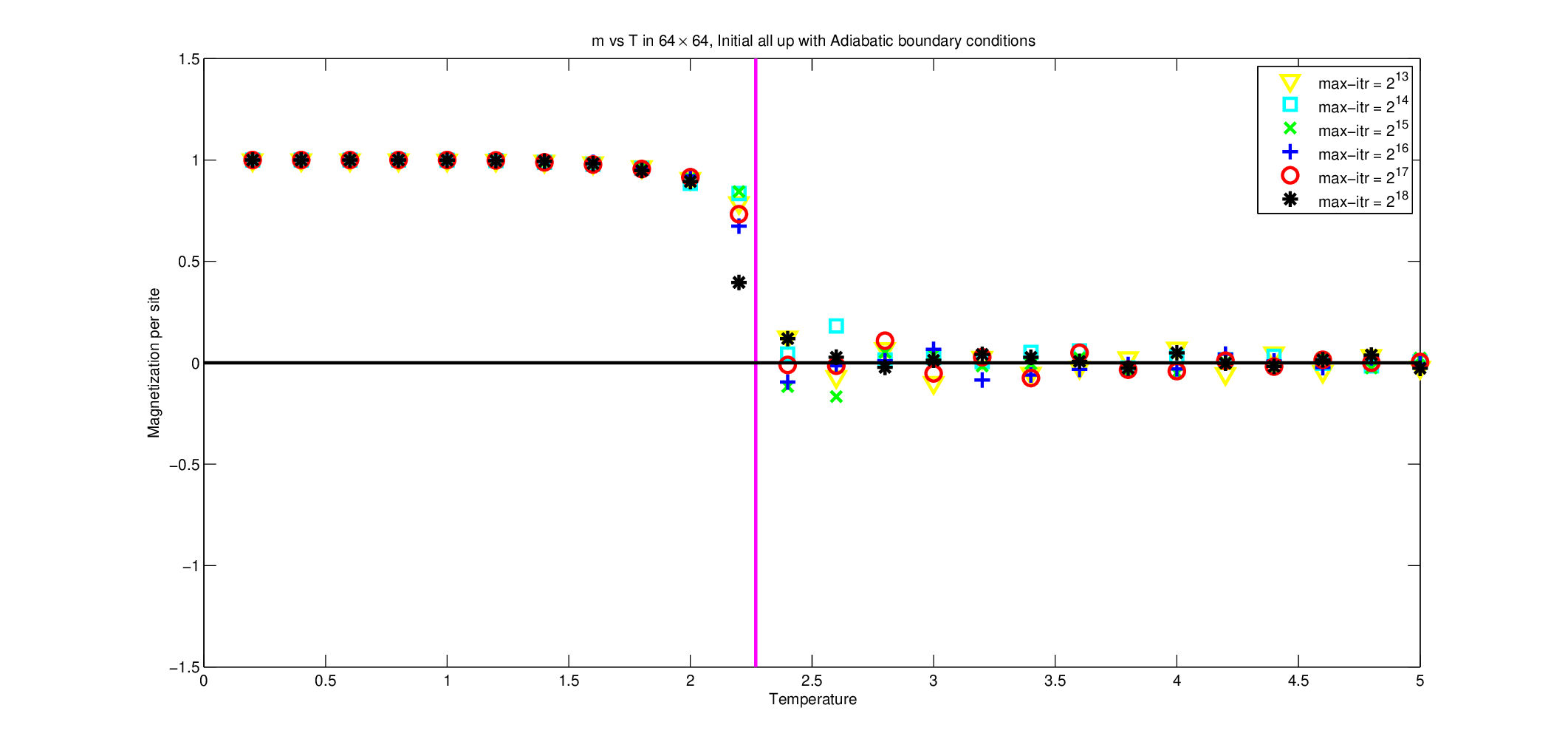}}
\vspace*{8pt}
\caption{Magnetisation per site vs temperature for
the comparison among different $t_{max}$ i.e., $2^{13}$,
$2^{14}$, $2^{15}$, $2^{16}$, $2^{17}$ and $2^{18}$
at all down spin initial condition and abc. Magenta line represents a
parallel line to magnetisation per site at $T=T_c$. \label{f2}}
\end{figure}

We have considered similar simulation procedure to find the optimal
$t_{max}$ with lattice size $64\times64$ for different
$t_{max}$ i.e., $2^{13}$, $2^{14}$, $2^{15}$, $2^{16}$, $2^{17}$
and $2^{18}$. One of the simulation result given in figure ~\ref{f2} shows
that $t_{max} = 2^{17}$ is the best.
%%%%%%%%%%%%%%%%%%%%%%%%%%%%%%%%%%%%%%%%%%%%%%%%%%%%%%%%%%%%%%%
%Figure 2
%%%%%%%%%%%%%%%%%%%%%%%%%%%%%%%%%%%%%%%%%%%%%%%%%%%%%%%%%%%

%%%%%%%%%%%%%%%%%%%%%%%%%%%%%%%%%%%%%%%%%%%%%%%%%%%%%%%%%%%%%
\subsection{Phase transition with pbc, abc, rbc, f1bc and f-1bc}
%%%%%%%%%%%%%%%%%%%%%%%%%%%%%%%%%%%%%%%%%%%%%%%%%%%%%%%%%%%%

In this simulation, we have considered $64\times64$ lattice size
and $t_{max} = 2^{17}$ with all bcs and the temperature
ranging from $0.1$ to $5.0$ with increment of $0.1$ unit.
In figure 3, we have plotted $\bf e$ vs $\bf T$, $\bf m$ vs $\bf T$, $\bf m$ vs $\bf e$,
$\bf \chi$ vs $\bf T$, $\bf C_v$ vs $\bf T$ and $\bf S$ vs $\bf T$ with initial condition
all up with all bcs. Figures 4 and 5 show similar plots
with initial condition all down and random spin configuration
respectively. We compare, the simulation result of magnetisation
with all five bcs with all three initial conditions with the
exact solution given by Onsager which are shown in $m$ vs $T$
graphs in figures 3(a), 4(a) and 5(a). Between temperature
$T = 2$ and $T = 2.5$, one finds that the energy gradually
increases, magnetisation gradually decreases to zero. The
susceptibility and specific heat also change, initially they
increase up to $T_c$ and then start decreasing as shown in
figures 3(d), 4(d), 5(d) and 3(e), 4(e), 5(e) respectively.
Entropy gradually increases between temperature $T = 2$ and $T = 2.5$
and the stays at maximum which are shown in figures 3(f), 4(f) and 5(f).
So, a phase transition is clearly visible in between $T = 2$ and
$T = 2.5$ with all three initial conditions with all five bcs.
In $m$ vs $e$ graphs shown in figures 3(c), 4(c) and 5(c),
the higher density states indicate four states. We find two
low temperature ground states around ($M = \pm1$, $E = -4$)
with all three initial conditions and all five bcs. The high
temperature phase is centered at ($M = 0$, $E = -1$) with all
three initial conditions with all five bcs. Then the other state is
around ($M = 0$, $E = -3.5$), which is a low-temperature metastable
states with all five bcs and  this happens only in case of random
initial condition. With initial condition all up spins and f-1bc,
one can find ground state at ($M = -1$, $E = -4$) and for other
bcs at ($M = +1$, $E = -4$) and with initial condition all down
spins and f1bc, one can find ground state at ($M = +1$, $E = -4$)
and for other bcs at ($M = -1$, $E = -4$).

%%%%%%%%%%%%%%%%%%%%%%%%%%%%%%%%%%%%%%%%%%%%%%%%%%%%%%%%%%%%%%%%%%%%%
\begin{figure}[ph]
\begin{tabular}{cc}
  \includegraphics[width=7cm]{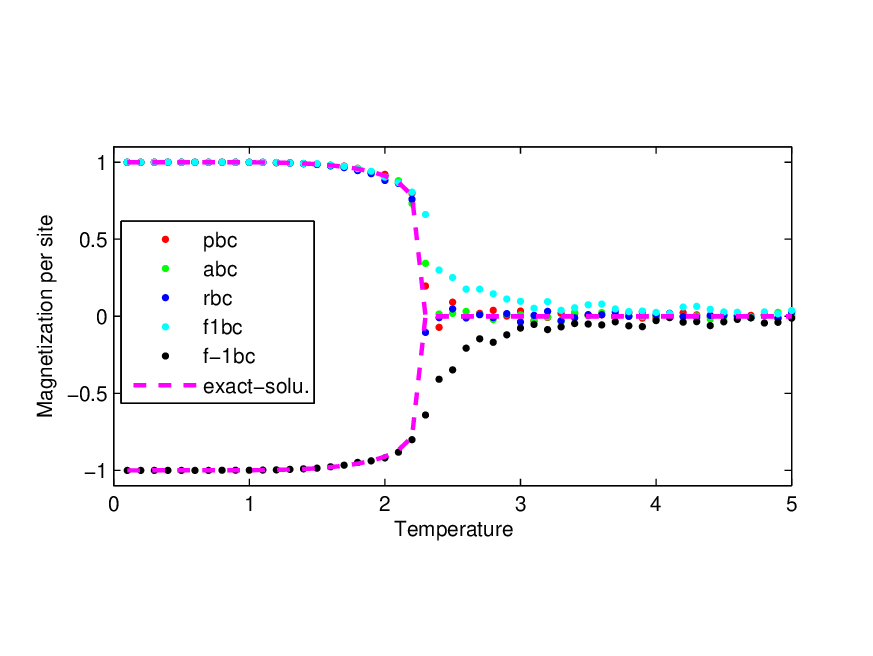} &   \includegraphics[width=7cm]{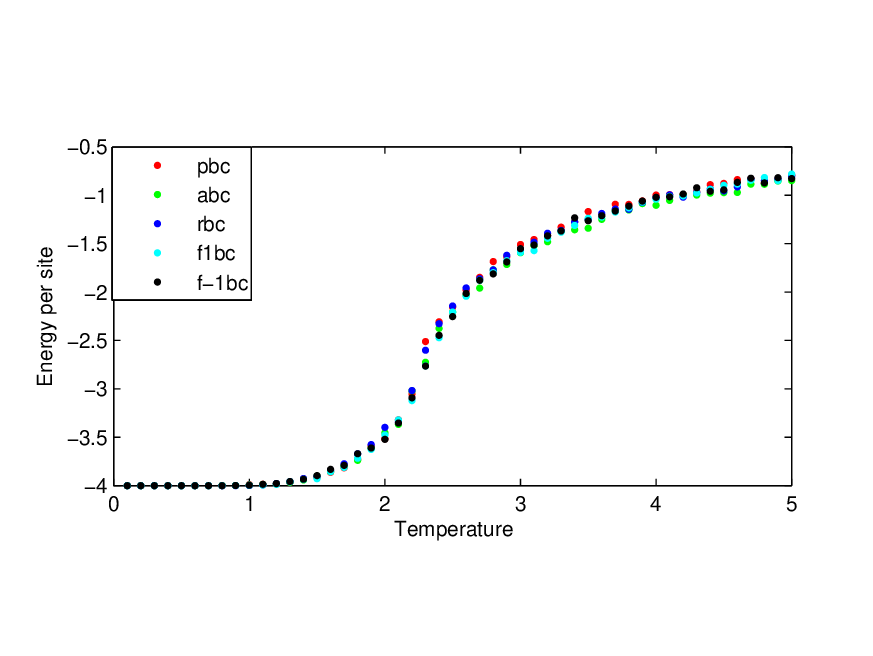}
  \\(a)  & (b)  \\
  \includegraphics[width=7cm]{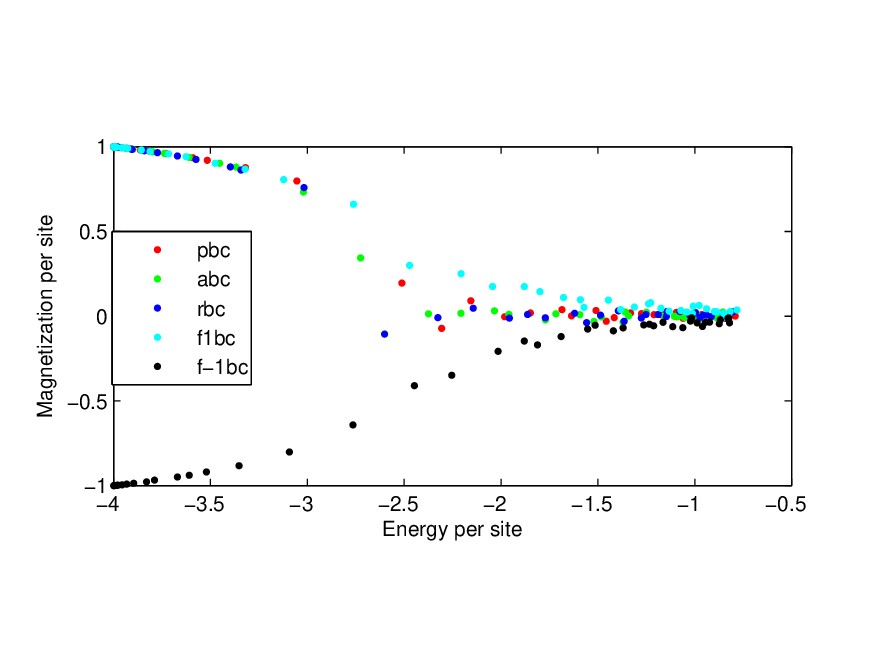} &   \includegraphics[width=7cm]{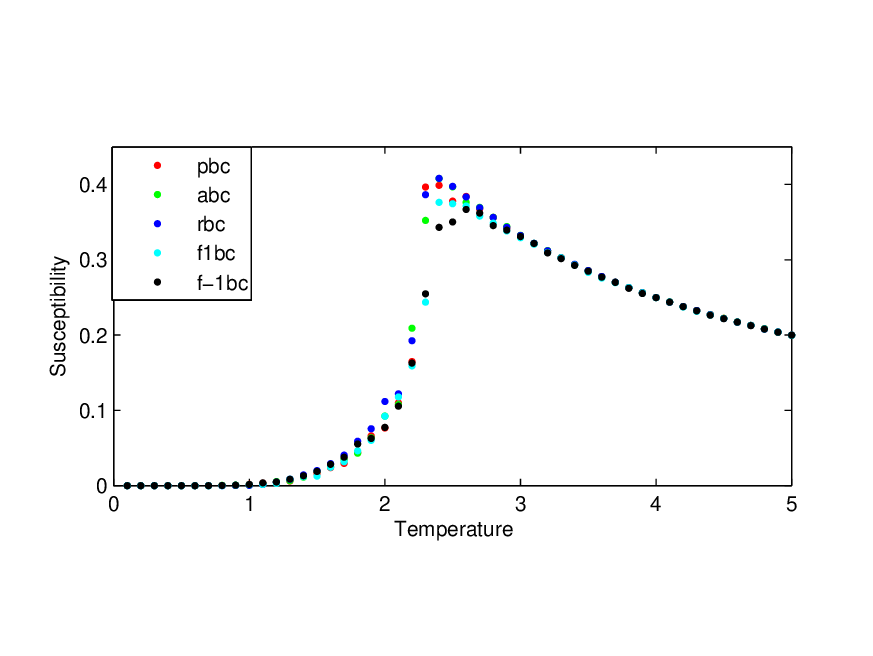}
  \\(c)  & (d)  \\
 \includegraphics[width=7cm]{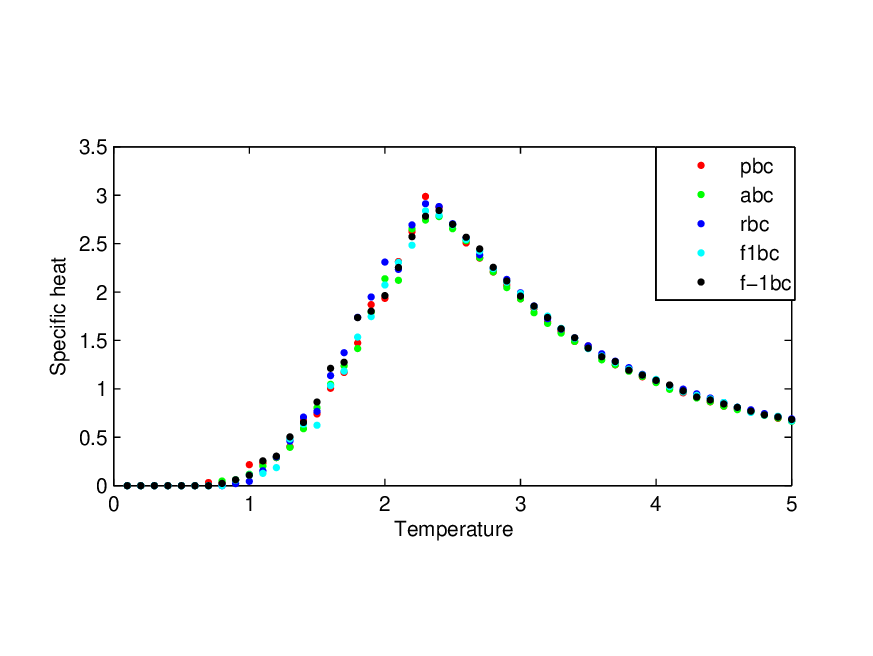} &   \includegraphics[width=7cm]{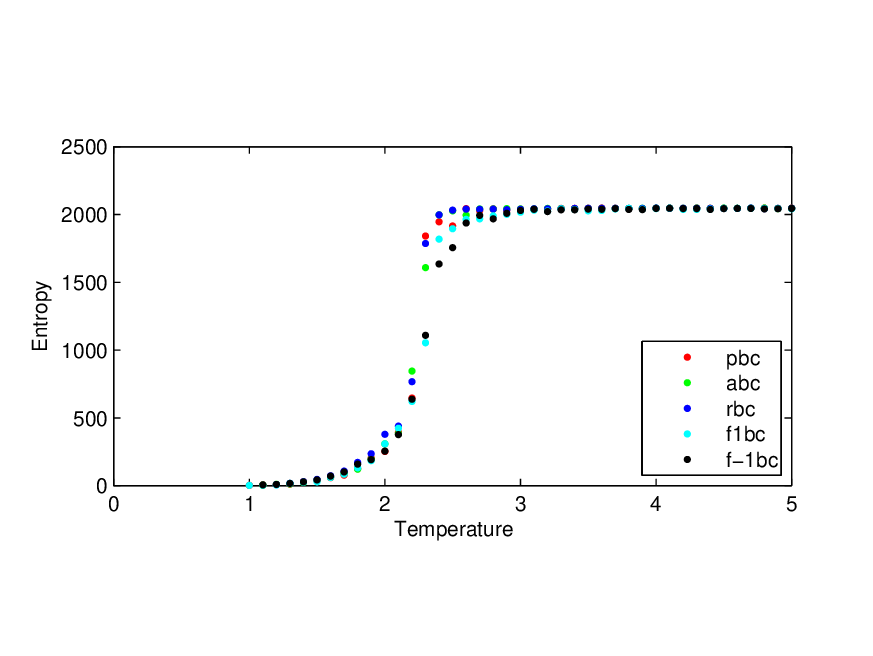}
  \\(e)  & (f)
\end{tabular}
\caption{(a) $m$ versus $T$ for all bcs, (b) $e$ versus $T$ for all bcs, (c) $m$ versus $e$ for all bcs, (d) $\chi$ versus $T$ for all bcs,
(e) $C_v$ versus $T$ for all bcs and (f) $S$ versus $T$ for all bcs. Initial condition with all up spin configuration for all five bcs. \label{f3}}
\end{figure}
%%%%%%%%%%%%%%%%%%%%%%%%%%%%%%%%%%%%%%%%%%%%%%%%%%%%%%%%%%%%%%%%%%%%%%%%%%%%%%%%%%%%%%%%%%%%%%%%%%%%%%%%
\begin{figure}[ph]
\begin{tabular}{ccc}
  \includegraphics[width=7cm]{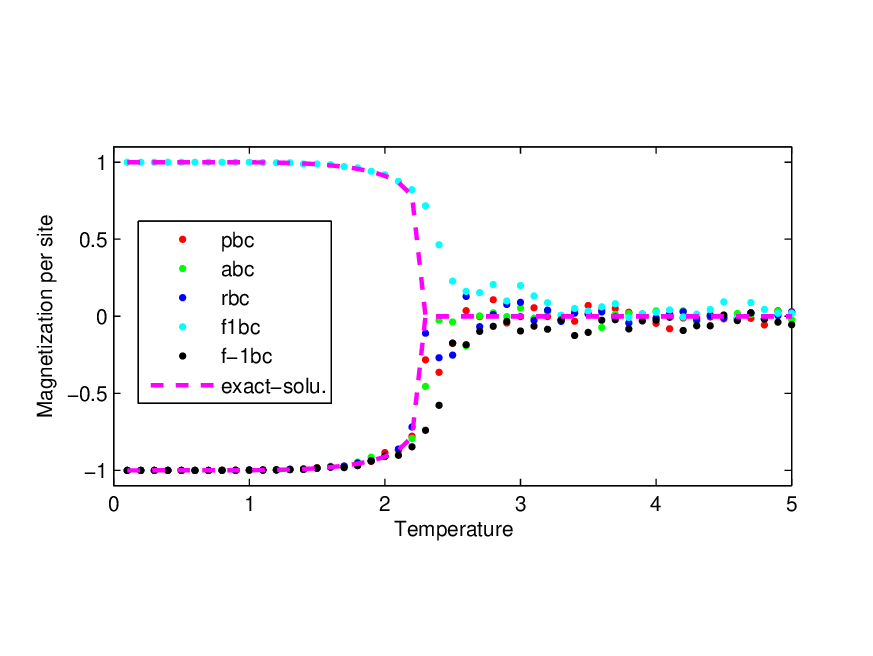} &   \includegraphics[width=7cm]{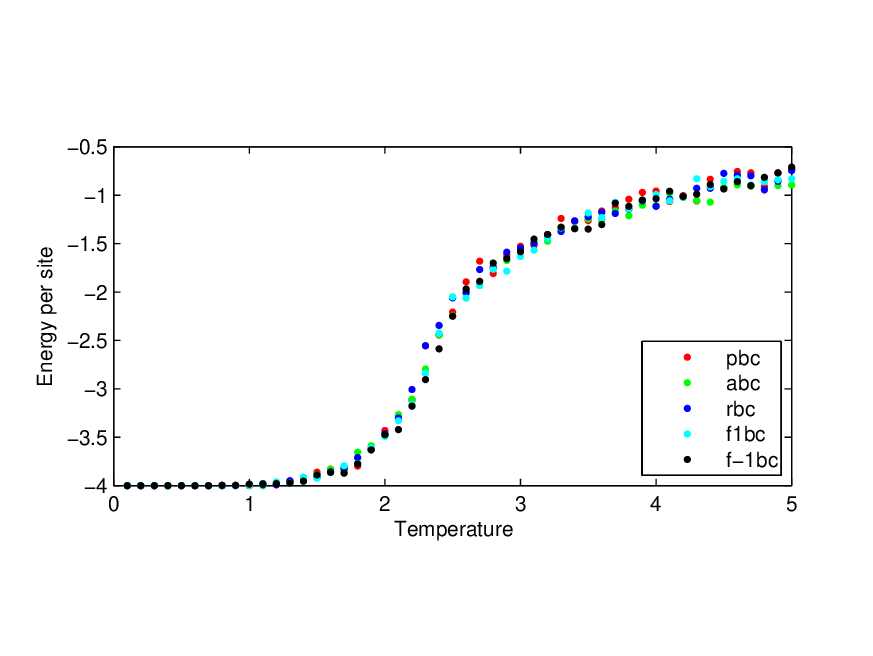}
  \\(a)  & (b)  \\
  \includegraphics[width=7cm]{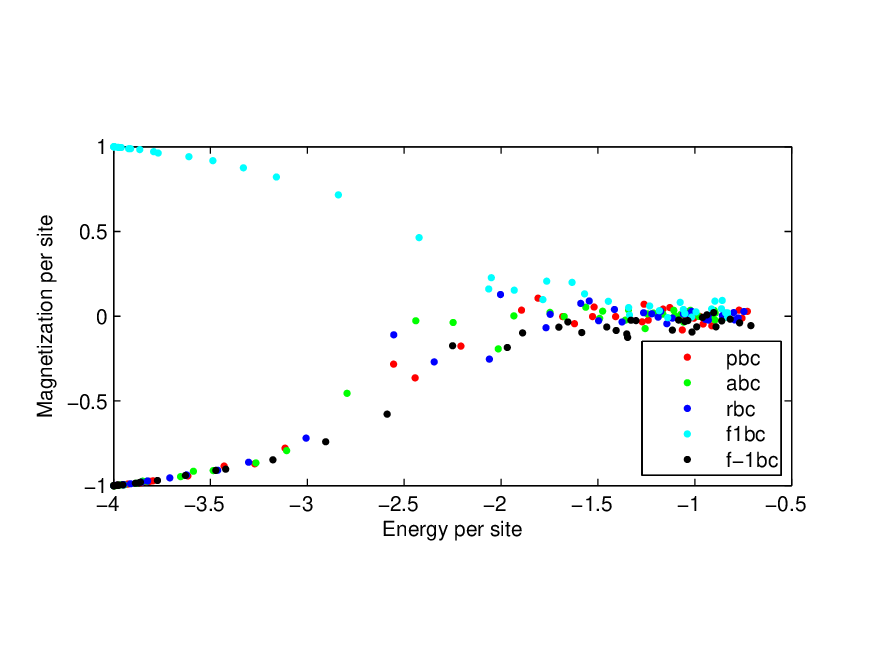} &   \includegraphics[width=7cm]{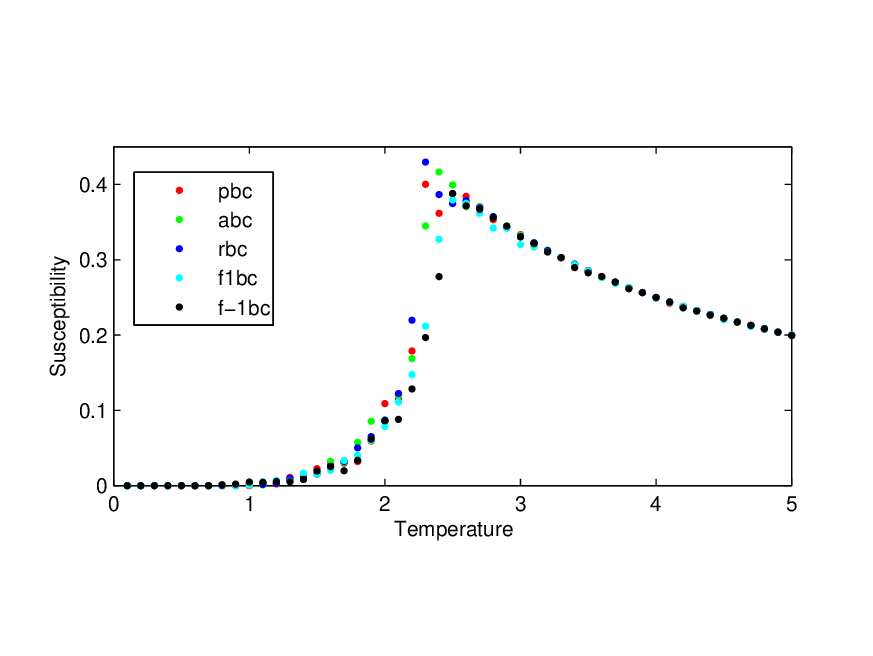}
  \\(c)  & (d)  \\
 \includegraphics[width=7cm]{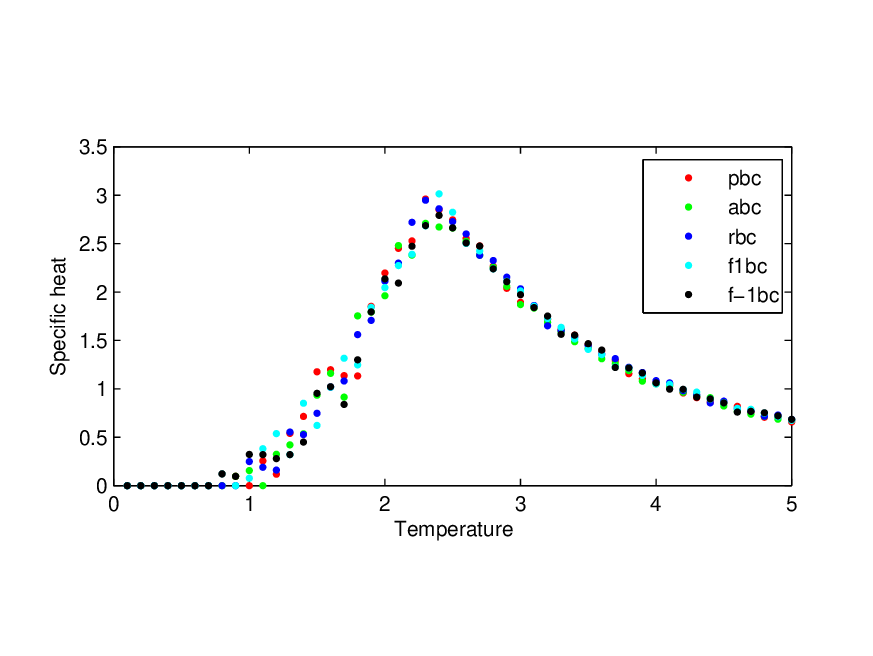} &   \includegraphics[width=7cm]{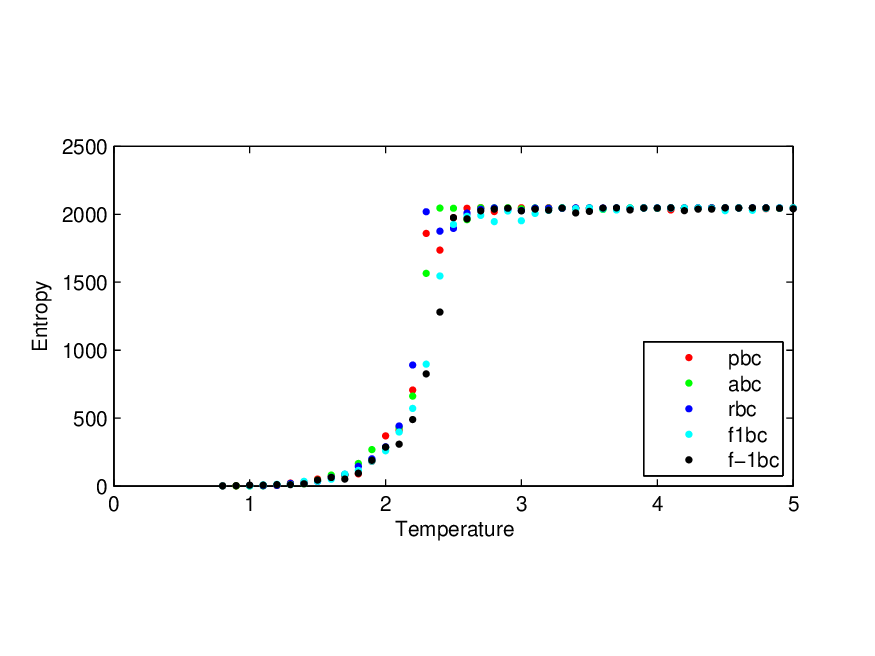}
  \\(e)  & (f)
\end{tabular}
\caption{(a) $m$ versus $T$ for all bcs, (b) $e$ versus $T$ for all bcs, (c) $m$ versus $e$ for all bcs, (d) $\chi$ versus $T$ for all bcs,
(e) $C_v$ versus $T$ for all bcs and (f) $S$ versus $T$ for all bcs. Initial condition with all down spin configuration for all five bcs. \label{f4}}
\end{figure}
%%%%%%%%%%%%%%%%%%%%%%%%%%%%%%%%%%%%%%%%%%%%%%%%%%%%%%%%%%%%%%%%%%%%%%%%%%%%%%%%%%%%%%%%%%%%%%%%%%%%%%%%%%%

\begin{figure}[ph]
\begin{tabular}{ccc}
  \includegraphics[width=7cm]{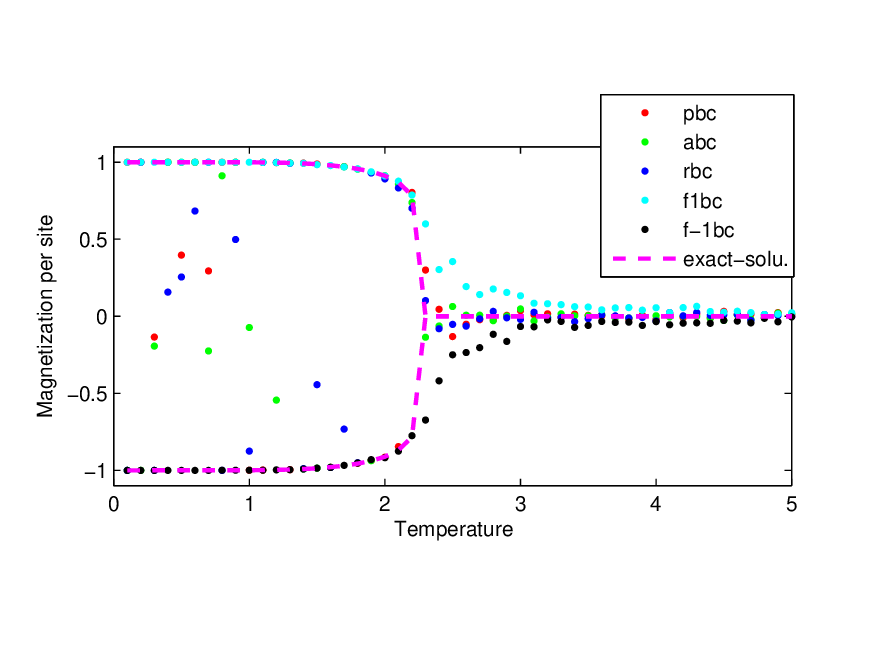} &   \includegraphics[width=7cm]{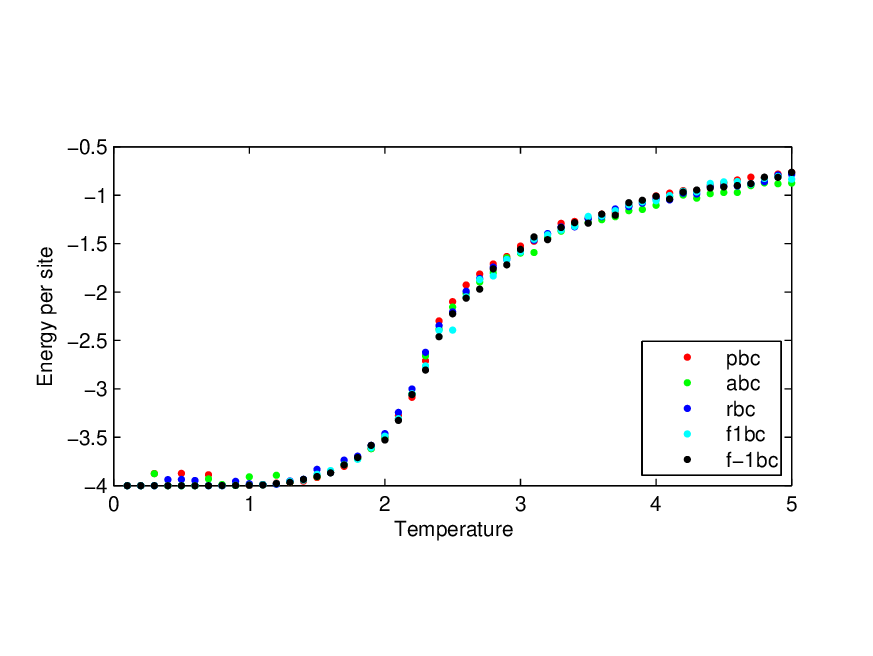}
  \\(a)  & (b)  \\
  \includegraphics[width=7cm]{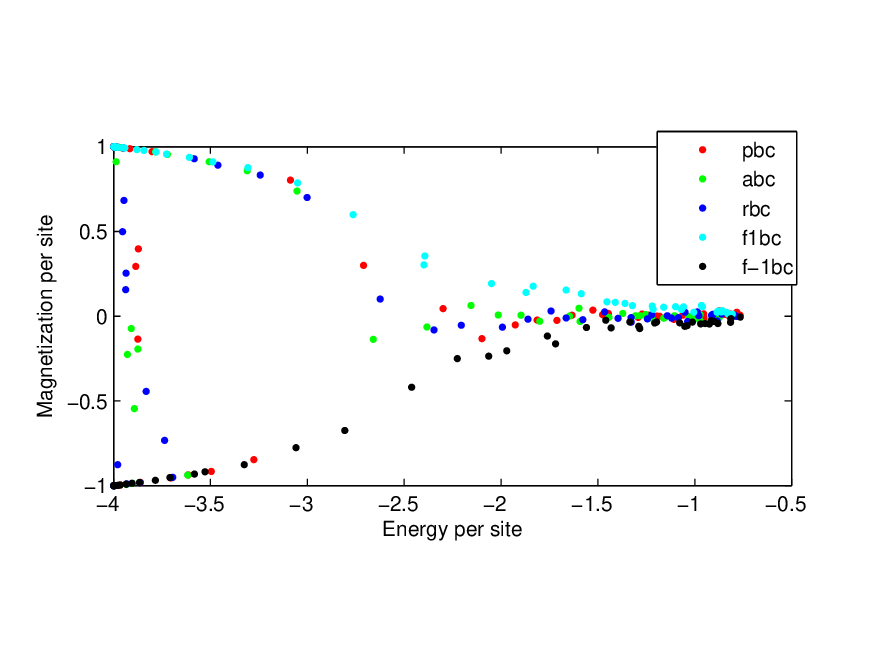} &   \includegraphics[width=7cm]{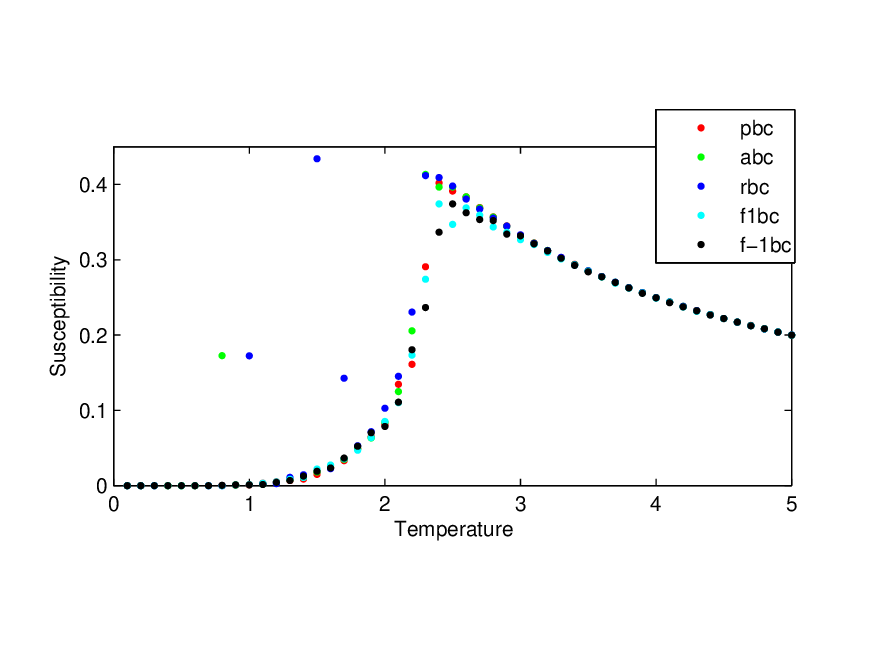}
  \\(c)  & (d)  \\
 \includegraphics[width=7cm]{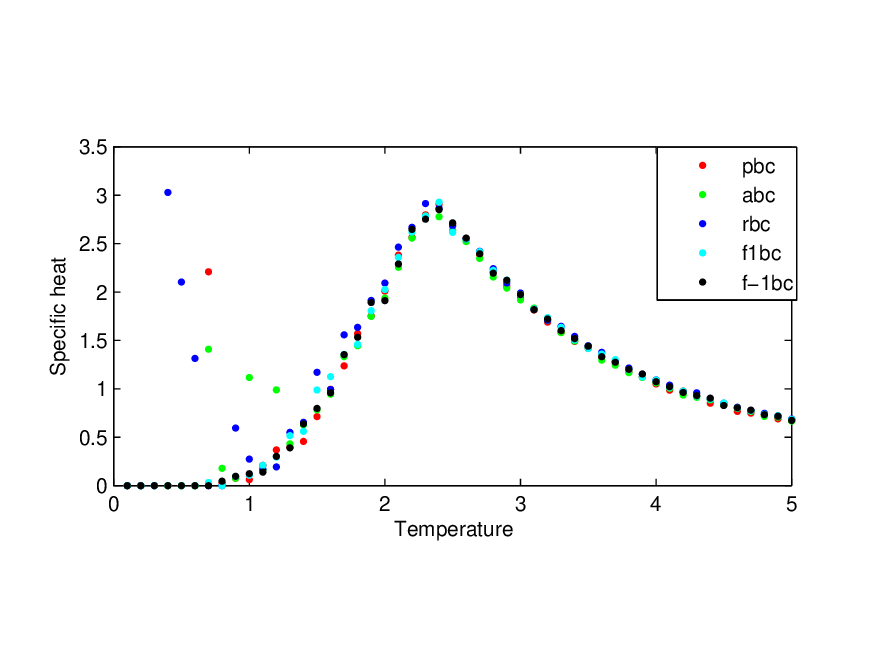} &   \includegraphics[width=7cm]{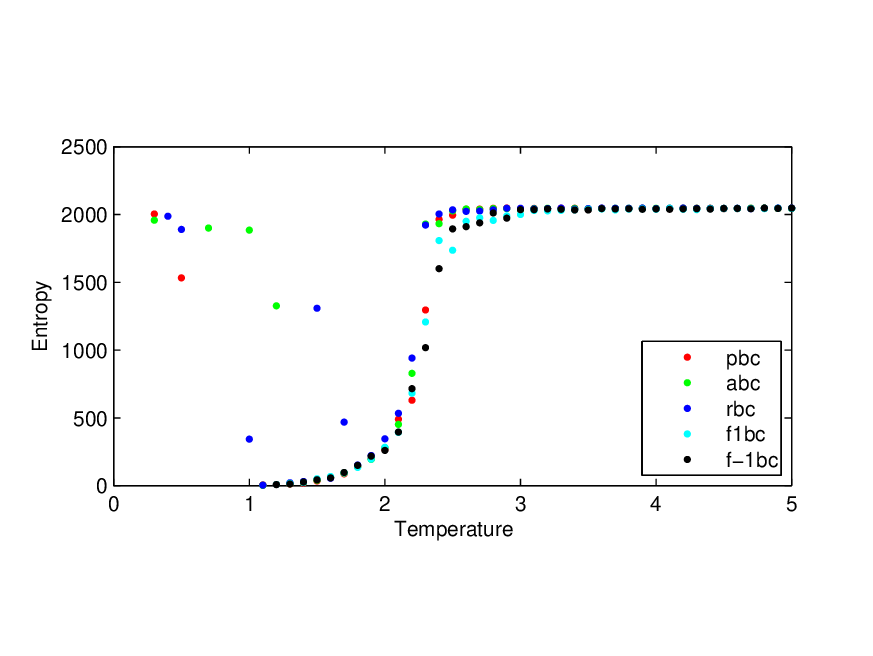}
  \\(e)  & (f)
\end{tabular}
\caption{(a) $m$ versus $T$ for all bcs, (b) $e$ versus $T$ for all bcs, (c) $m$ versus $e$ for all bcs, (d) $\chi$ versus $T$ for all bcs,
(e) $C_v$ versus $T$ for all bcs and (f) $S$ versus $T$ for all bcs. Initial condition with random spin configuration for all five bcs. \label{f5}}
\end{figure}
%%%%%%%%%%%%%%%%%%%%%%%%%%%%%%%%%%%%%%%%%%%%%%%%%%%%%%%%%%%%%%%%%%%%%%%%%%%%%%%%%%%%%%%%%%%%%%%%%%%%%

%%%%%%%%%%%%%%%%%%%%%%%%%%%%%%%%%%%%%%%%%%%%%%%%%%%%%%%%%%%%%
\section{Comparison among Boundary Conditions}
\label{sec:comparision}
%%%%%%%%%%%%%%%%%%%%%%%%%%%%%%%%%%%%%%%%%%%%%%%%%%%%%%%%%%%%%
Starting with three different initial conditions, from figure 2,
one finds that the magnetisation meet the zero line
after $T>T_c$ differently which is closer
to the exact solution in all bcs.

With one simulation for all bcs, it is not possible to predict
which bc is closer to $T_c$. So, we analyse the
points for magnetisation in the range  $-0.1 \le m \le 0.1$ and
$-0.2 \le m \le 0.2$ which are close to the
zero line of magnetization (where magnetisation is zero)
after $T>T_c$. We call such points as converging points.

For the above purpose, we have taken different lattice
sizes ranging from $5\times5$ to $60\times60$ with increment
of $5$;  from $60\times60$ to $100\times100$ with increment
of $10$ and temperature ranging from $0.1$ to $5.0$
with small increment of $0.05$ units. We have considered $t_{max} = 2^{15}$
for lattice size $\le 30\times30$ and $t_{max} = 2^{17}$ for lattice size $>30\times30$.
Figure 6 shows converging
points in the above mentioned range of $m$. We have counted the
number of converging points as defined above. Their average percentage
have calculated by taking average of ten simulations result
of each bc with three different initial conditions.
%%%%%%%%%%%%%%%%%%%%%%%%%%%%%%%%%%%%%%%%%%%%%%%%%%%%%%%%%%%%%%%%%%%%
\begin{figure}[ph]
\centerline{\includegraphics[width=13.0cm]{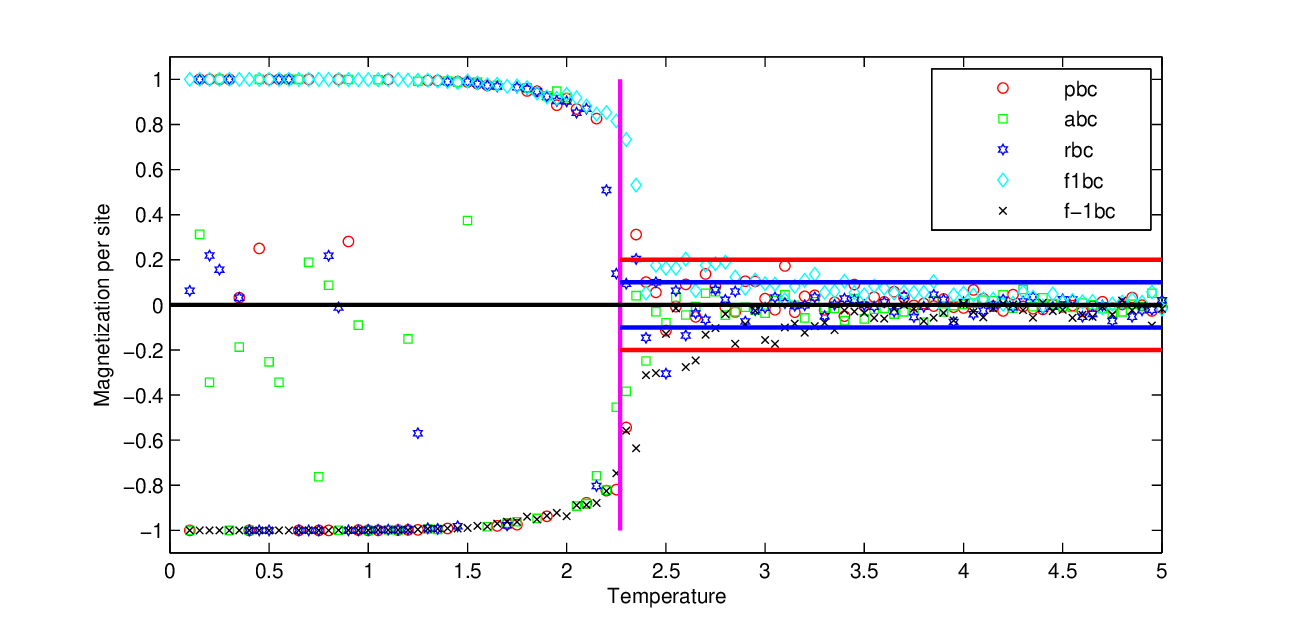}}
\vspace*{8pt}
\caption{Converging points of bcs for $60\times60$ lattice size
after $T>T_c$. Red lines are for magnetisation $m = +0.2$ and $m = -0.2$
and blue lines are for magnetisation $m = +0.1$ and $m = -0.1$ with the
initial condition of random spin configuration. Magenta line represents a
parallel line to magnetisation per site at $T=T_c$. \label{f6}}
\end{figure}
%%%%%%%%%%%%%%%%%%%%%%%%%%%%%%%%%%%%%%%%%%%%%%%%%%%%%%%%%%%
Lattice $size \le 30\times30$ with pbc, abc and rbc, one
finds more convergent points in both cases
$-0.1 \le m \le 0.1$ and $-0.2 \le m \le 0.2$ than f1bc
and f-1bc but among the three rbc shows more converging
points in all three initial conditions. With lattice size
between $30\times30$ and $70\times70$ with pbc, abc, rbc,
one finds more convergent points in both the cases
$-0.1 \le m \le 0.1$ and $-0.2 \le m \le 0.2$ in comparison
to f1bc and f-1bc with random initial case. For all initial
cases with $-0.1 \le m \le 0.1$, the result is better for pbc,
abc and rbc. For $-0.2 \le m \le 0.2$, the results are better
for f1bc with initial condition all down spins and f-1bc with
initial condition all up spins. For lattice size greater than
$70\times70$ and less than equal to $100\times100$ with initial
conditions of all up spins and f-1bc and initial condition of
all down spins and f1bc, one observe more converging points
in both cases $-0.1 \le m \le 0.1$ and $-0.2 \le m \le 0.2$.
For random initial condition with pbc, rbc and abc, one finds
more converging points in both the range of magnetisation per site.

%%%%%%%%%%%%%%%%%%%%%%%%%%%%%%%%%%%%%%%%%%%%%%%%%%%%%%%%%%%%
\section{Conclusion}
\label{sec:conclusion}
%%%%%%%%%%%%%%%%%%%%%%%%%%%%%%%%%%%%%%%%%%%%%%%%%%%%%%%%%%%
We have observed a second order phase transition with respect to all
boundary conditions considering all the initial conditions around
the critical temperature $T_c$.
This implies that with the different initial conditions
on different lattice sizes $\le 30\times30$, one can take
care of boundary spins by not only pbc but also by abc and rbc.
Further in our analysis, rbc shows more converging
points than pbc and rbc for lattice size $\le 30\times30$.
For lattice size greater than $70\times70$, f1bc and
f-1bc are better suited to use than other bcs in case of initial spin
configuration with all up or all down spins. It is also observed
that, in case of random initial spin configuration, it is better to use either pbc,
abc or rbc when the lattice size is $\le 100\times100$.
It will be interesting to study the behaviour of all five bcs for lattice
$size > 100^2$ with all initial conditions. Further, also study the
phase transition with different values of $t_{max}$ for different lattice sizes.
From the simulation point of view, our method takes lesser time than the Metropolis
algorithm \cite{5}. This observation is expected to find
the approximate values of the critical exponents more accurately. Here, one can
reduce the simulation time by generating random numbers using deterministic CA.
%%%%%%%%%%%%%%%%%%%%%%%%%%%%%%%%%%%%%%%%%%%%%%%%%%%%%%%%%%

%%%%%%%%%%%%%%%%%%%%%%%%%%%%%%%%%%%%%%%%%%%%%%%%%%%%%%%%%%%%%%%%%%%%%%
\subsection*{Acknowledgement}
Work of JM is supported by UGC under the Basic Scientific Research
(BSR) scheme. We thank K. Maharana and S. Pattanayak
for useful discussions.

%%%%%%%%%%%%%%%%%%%%%%%%%%%%%%%%%%%%%%%%%%%%%%%%%%%%%%%%%%%%%%%%%%%%%%

\end{document}